\documentclass[12pt,preprint]{aastex}

\def\mv{M_V}
\def\msun{$M_{\odot}$}

\def\Te{T_{\rm eff}}
\def\logg{\log g}
\def\gta{\lower 0.5ex\hbox{$\buildrel > \over \sim\ $}} 
\def\lta{\lower 0.5ex\hbox{$\buildrel < \over \sim\ $}} 

\tighten

\begin{document}

\title{DISCOVERY OF A COOL, MASSIVE, AND METAL-RICH DAZ WHITE DWARF}

\author{A. Gianninas, P. Dufour and P. Bergeron }
\affil{D\'epartement de Physique, Universit\'e de Montr\'eal, C.P.~6128, 
Succ.~Centre-Ville, Montr\'eal, Qu\'ebec, Canada, H3C 3J7.}
\email{gianninas, dufourpa, bergeron@astro.umontreal.ca}

\begin{abstract}

We report the discovery of a new metal-rich DAZ white dwarf, GD
362. High signal-to-noise optical spectroscopy reveals the presence of
spectral lines from hydrogen as well as Ca~\textsc{i}, Ca~\textsc{ii},
Mg~\textsc{i}, and Fe~\textsc{i}.  A detailed model atmosphere
analysis of this star yields an effective temperature of
$\Te=9740\pm50$~K, a surface gravity of $\logg=9.12 \pm 0.07$, and
photospheric abundances of $\log{\rm Ca/H}=-5.2\pm0.1$, $\log{\rm
Mg/H}=-4.8\pm0.1$, and $\log{\rm Fe/H}=-4.5\pm0.1$. White dwarf
cooling models are used to derive a mass of 1.24
\msun\ for GD 362, making it the most massive and metal-rich DAZ
uncovered to date. The problems related to the presence of such large
metal abundances in a nearby ($d\sim 25$~pc) white dwarf in terms of
an accretion scenario are briefly discussed.

\end{abstract}


\section{INTRODUCTION}

The most common type of white dwarfs are those that show hydrogen
absorption lines in their optical spectra, the DA stars. A fraction of
these, the so-called DAZ stars, also display absorption features from
heavier elements, most notably the Ca~\textsc{ii} H and K lines. A
recent spectroscopic survey at high dispersion of about 120 cool
($\Te<10,000$~K) DA stars by \citet{zuckerman03} reveals that $\sim
25$\% of the objects in their sample show metallic features. The DAZ
phenomenon thus seems much more common than previously believed
\citep{billeres97}.  The origin of these metals in the atmospheres of
DAZ stars is still poorly understood, however. Indeed, as most DAZ
stars are comparatively cool and old, it is expected that all elements
heavier than hydrogen would have settled rather quickly at the bottom
of the atmosphere due to the strong gravitational field present in
these stars. Therefore, any metals found in the atmosphere must be
provided by an exterior source, such as the interstellar medium or
even comets. We refer the reader to \citet{zuckerman03} for a more
detailed discussion of the various theories that have been put forth
to account for the presence of metals in DAZ stars.

We have recently undertaken a spectroscopic survey of white dwarfs
drawn from the catalog of \citet{mccook99}. Our aim is to derive the
atmospheric parameters for all stars, as well as to confirm the
spectroscopic classifications given in the catalog using modern CCD
spectroscopy. As part of this survey, we have discovered that GD 362
(WD 1729+371, PG) is a unique DAZ white dwarf, with metallic features
stronger than in any other member of its class, the Ca~\textsc{ii} H
and K lines in particular. GD 362 had initially been classified as a
DA white dwarf by \citet{greenstein80} with a note that it could
possibly constitute a composite spectrum or a dK subdwarf in a common
proper motion binary system. It was later rediscovered in the
Palomar-Green survey \citep{green86} where it was classified
\textit{sd}, which does not imply a subdwarf type, but rather ``a
lower signal-to-noise observation in which two or three Balmer
absorption lines of moderate gravity are visible''

We present in this {\it Letter} our spectroscopic data for GD 362 in
which, besides the usual hydrogen Balmer lines, we have
identified spectral lines of Ca~\textsc{i}, Ca~\textsc{ii},
Mg~\textsc{i} and Fe~\textsc{i}. By fitting the optical spectrum with
a grid of synthetic spectra appropriate for these stars, we determine the 
effective temperature and surface gravity of GD 362, as well as the 
abundances of all the elements observed in this star.

\section{FITTING PROCEDURE AND SYNTHETIC SPECTRA}

\subsection{Observations}

The high signal-to-noise (S/N$\sim 110$ per pixel) optical spectrum of
GD 362 was obtained over the course of 2 nights using the Steward
Observatory 2.3 m telescope equipped with the Boller \& Chivens
spectrograph.  The 4.5 arcsec slit together with the 600 l/mm grating
blazed at $\lambda3568$ in first order provided a spectral coverage
from about 3000 to 5250~\AA\ at a resolution of $\sim$ 6~\AA\
FWHM. The spectra were reduced using standard IRAF packages. The
spectrum is displayed in Figure
\ref{fg:f1} where the major spectral features are identified, 
namely Ca~\textsc{i}, Ca~\textsc{ii}, Mg~\textsc{i}, and Fe~\textsc{i} lines.  
We also show for comparison the optical spectrum of G74-7,
discovered by \citet{lacombe83}, which was the only DAZ star known 
until the more recent discoveries of G238-44
\citep{holberg97}, G29-38 \citep{koester97}, and those reported 
by \citet{zuckerman03}. The comparison is most striking when we
examine in both spectra the relative strengths of the H and K lines of
Ca~\textsc{ii} at 3933.66~\AA\ and 3968.67~\AA, respectively.  It is
worth noting that the region blueward of $\sim3800$~\AA\ is heavily
blanketed by Fe \textsc{i} lines.

\subsection{Fitting Procedure}

The technique for fitting the spectrum is similar to that
described at length by \citet{liebert04}. First the continuum
used to normalize the spectrum is set by fitting a model spectrum 
to the data, allowing for a possible wavelength shift, a zero point offset, 
and higher order terms in $\lambda$ (up to $\lambda^{3}$). The purpose
of this procedure is simply to obtain a function that defines best 
the continuum of the observed spectrum. Once the spectrum has been normalized properly, 
the atmospheric parameters and chemical abundances are obtained from a 
grid of model spectra, convolved with a Gaussian profile at 6~\AA\ FWHM,
using the nonlinear least-squares method of Levenberg-Marquardt 
\citep{press86}. This minimization technique also provides formal 
uncertainties of the fitted parameters through the covariance matrix
\citep[see][for details]{BSL}.

\subsection{Synthetic Spectra}

For the computation of our synthetic spectra, we adopt a grid of LTE
model atmospheres with a pure hydrogen composition similar to that
described in \citet{liebert04}. We thus make the assumption that the
metals have no effect on the atmospheric structure and include them
only in the calculation of the emergent fluxes; this assumption will
be tested subsequently. Atomic partition functions have been
calculated by summing explicitly over the bound states taken directly
from TOPBASE. Central wavelengths of the transitions, $gf$ values,
energy levels, and damping constants (radiative, Stark, and van der
Waals) have been extracted from the GFALL linelist of Kurucz
(http://kurucz.harvard.edu/LINELISTS.html).

Since our preliminary analysis of GD 362 \citep{gianninas05} revealed
that it was a high surface gravity DAZ star, we restrict our model
grid calculations to high values of $\logg$. We first begin with the
determination of the calcium abundance since calcium produces the most
prominent features in the spectrum.  Our model grid covers a range of
$-5.7 \le \log{\rm Ca/H}\le -4.8$ in steps of 0.3 dex, effective
temperatures in the range $9000\le\Te\le10,400$~K in steps of 200 K,
and surface gravities in the range $8.75\le\logg\le9.50$ in steps of
0.25 dex. With the calcium abundance fixed, we proceed to determine in
a similar way the magnesium and iron abundances in turn, using spectra
with magnesium abundances in the range $-5.6 \le \log$ Mg/H $\le -4.4$
in steps of 0.4 dex, and iron abundances in the range $-5.0 \le \log$
Fe/H $\le -4.4$ in steps of 0.2 dex. At each step, the determination
of the effective temperature and surface gravity is refined.

\subsection{Atmospheric Parameter and Abundance Determinations}

Before undertaking our analysis of GD 362, we wished to test our grid
of synthetic spectra, and more importantly our fitting procedure. With
this in mind, we fitted the optical spectrum of G74-7 analyzed
by \citet{billeres97} who derived $\Te=7260 \pm 40$~K, $\log g=8.03 \pm 0.07$, and 
$\log {\rm Ca/H}=-8.8 \pm 0.1$. Our own analysis of the same spectrum
yields $\Te=7300 \pm 60$~K, $\log g=8.02 \pm 0.11$, and 
$\log {\rm Ca/H}=-8.9 \pm 0.3$, in excellent agreement with the results of
Bill\`eres et al. 

Our best fit to the spectrum of GD 362 is displayed in Figure
\ref{fg:f2}. The atmospheric parameters and corresponding uncertainties are
$\Te=9740\pm50$~K, $\logg=9.12\pm 0.07$, $\log {\rm Ca/H}=-5.2\pm0.1$,
$\log{\rm Mg/H}=-4.8\pm0.1$, and $\log{\rm Fe/H}=-4.5\pm0.1$ . We are
able to obtain a very good fit to the observed spectrum albeit not a
perfect one. First, we notice that although the H and K lines of
Ca~\textsc{ii} are extremely well reproduced, the Ca~\textsc{i} lines at
4227 and 4585~\AA\ are predicted
too strong. We believe that this is caused by an incorrect
Ca~\textsc{ii}/Ca~\textsc{i} ratio which could be due either to an
erroneous electron density or partition function values obtained from
TOPBASE. Second, the Fe~\textsc{i} lines in the bluest portion of the spectrum
are not reproduced perfectly. Some of the atomic data for iron is
derived from theoretical calculations and is more uncertain than
values determined experimentally. We therefore suggest that our
derived iron abundance is at worse an upper limit.

A posteriori, we computed a blanketed model atmosphere in which we 
took into account the presence of metals at our derived abundances. 
The synthetic spectrum calculated from this model
is identical to that computed here from our pure 
hydrogen models, validating our initial assumption that the metals play a
negligeable role in the determination of the atmospheric structure.

\section{DISCUSSION}

Using the evolutionary models of \citet{fontaine01} for DA white
dwarfs, we derive a mass of 1.24 \msun.  Our derived atmospheric
parameters thus make GD 362 the most metal-rich and massive DAZ white
dwarf studied thus far, and somewhat hotter than the majority of DAZ
stars discovered by \citet{zuckerman03}. In Figure \ref{fg:f3}, we
show the calcium abundance as a function of effective temperature for
the white dwarfs from \citet{zuckerman03}. We see from its location in
Figure \ref{fg:f3} how unique GD 362 truly is. In fact, if we compare
with the values listed in Table 4 of
\citet{zuckerman03}, the iron and magnesium abundances are nearly
solar! 

A comparison of our calcium abundance determination for GD 362 with
those of other DAZ stars from \citet{zuckerman03} with similar
effective temperatures reveals that the abundance observed in GD 362
is at least a factor of $10^{3}$ higher (see Fig.~\ref{fg:f3}). Since
the mass of GD 362 is also higher than average, we need to verify
whether this is the reason for the unusual observed photospheric
abundances. To do so, we compare the expected steady-state calcium
abundance in a massive star at 1.2 \msun\ to that of a normal star
with a mass of 0.6 \msun.  The steady-state abundance, $X_{\rm ss}$,
is given by equation (4) of \citet{dupuis93},

\begin{equation}
X_{\rm ss}=\frac{\theta \dot{M}}{\Delta M_{\rm cz}}\ \ ,
\end{equation}

\noindent where $\theta$ is the diffusion time scale of calcium at the 
bottom of the hydrogen convection zone, $\dot{M}$ is the accretion
rate of that element, and $\Delta M_{\rm cz}$ is the mass of the
convection zone. If we assume the Bondi-Hoyle accretion rate, then
$\dot{M}$ is proportional to $M^2$, where $M$ is the mass of the white
dwarf.  Then the expected relative abundance of calcium between a 1.2
and a 0.6 \msun\ white dwarf can be expressed as

\begin{equation}
\frac{X_{1.2}}{X_{0.6}} \propto \frac{\theta_{1.2}}{\theta_{0.6}} \ \ \frac{M_{1.2}^{2}}{M_{0.6}^{2}} \ \ \frac{\Delta M_{{\rm cz}, 0.6}}{\Delta M_{{\rm cz}, 1.2}}\ \ ,
\end{equation}

\noindent where we assume that both stars are moving with the same 
velocity through an interstellar medium of similar density.
Using values of $\theta$ and $\Delta M_{\rm cz}$ generously
provided to us by G.~Fontaine (private communication), we obtain that
 
\begin{displaymath}
\frac{X_{1.2}}{X_{0.6}} \propto \frac{1}{36} \cdot 4 \cdot 8.5 \approx 1\ \ .
\end{displaymath}

\noindent We see that even though a more massive star accretes at a higher rate and
has a thinner convection zone --- implying a smaller dilution factor,
the diffusion time scale at the bottom of the hydrogen convection zone
is so much shorter that comparable abundances are predicted.

Additionally, the accretion model suffers from the fact there is
little interstellar matter within 100 pc from the Sun, as
\citet{aanestad93} aptly pointed out. The distance to GD 362
can be estimated in two ways. First, the $\Te$ and $\logg$ values
inferred for GD 362 together with the photometric calibration of
\citet{bwb95} yield an absolute visual magnitude of $\mv=14.31$ (or
$M_B=14.48$), which combined with the photographic magnitude of
$B_{\rm ph}=16.15$ taken from \citet{green86}, yields a distance
estimate of $\sim 22$~pc. Alternatively, we can use the relation
between the monochromatic fluxes from our optical spectrum,
$f_{\lambda}$, and the theoretical Eddington fluxes, $H_{\lambda}$,
given by

\begin{equation}
f_{\lambda}=4\pi (R/D)^{2}H_{\lambda}\ \ ,
\end{equation}

\noindent where $R$ and $D$ are the radius of the star and its distance 
from Earth, respectively, to obtain an independent distance estimate 
of $\sim 26$~pc, consistent with our previous estimate. In either case, GD
362 seems to be well within the local bubble and far from any high
concentrations of interstellar matter. We are therefore at odds to
explain the very high metal abundances observed.

Another possibility would be that GD 362 is an unresolved degenerate
binary composed of a normal DA star and a helium-rich DZ star of
comparable luminosities. If this were the case, however, the intrinsic
absorption features of each star would be even stronger than observed
here, because of the line dilution factor, and the effective
temperature inferred from the Balmer lines would thus be grossly {\it
underestimated} (Balmer lines get stronger with increasing $\Te$ in
this temperature range). In such unresolved systems, the slope of the
theoretical spectrum is incompatible with that of the observed spectrum
\citep[see, e.g.,][]{bergeron90}. We see on the contrary from Figure
\ref{fg:f2} that the slope of the theoretical
spectrum is in perfect agreement with the observed slope of GD 362,
consistent with the presence of a single star.

It is perhaps not too suprising that GD 362 had not been classified as
a DAZ white dwarf by \citet{greenstein80} or \citet{green86}.  The
very strong Ca~\textsc{ii} H and K lines seen on their photographic
spectrum, in addition to the narrow Balmer lines, have led these
investigators to classify GD 362 as composite or as a lower gravity
object. This raises the interesting possibility that many such objects
exist and have been overlooked both in the past and in modern surveys
such as the Sloan Digital Sky Survey.

We thank Gilles Fontaine and Pierre Brassard for useful discussions.
We would also like to thank the director and staff of Steward
Observatory for the use of their facilities, as well as the referee
for useful suggestions. This work was supported in part by the NSERC
Canada and by the FQRNT (Qu\'ebec).

\clearpage

\figcaption[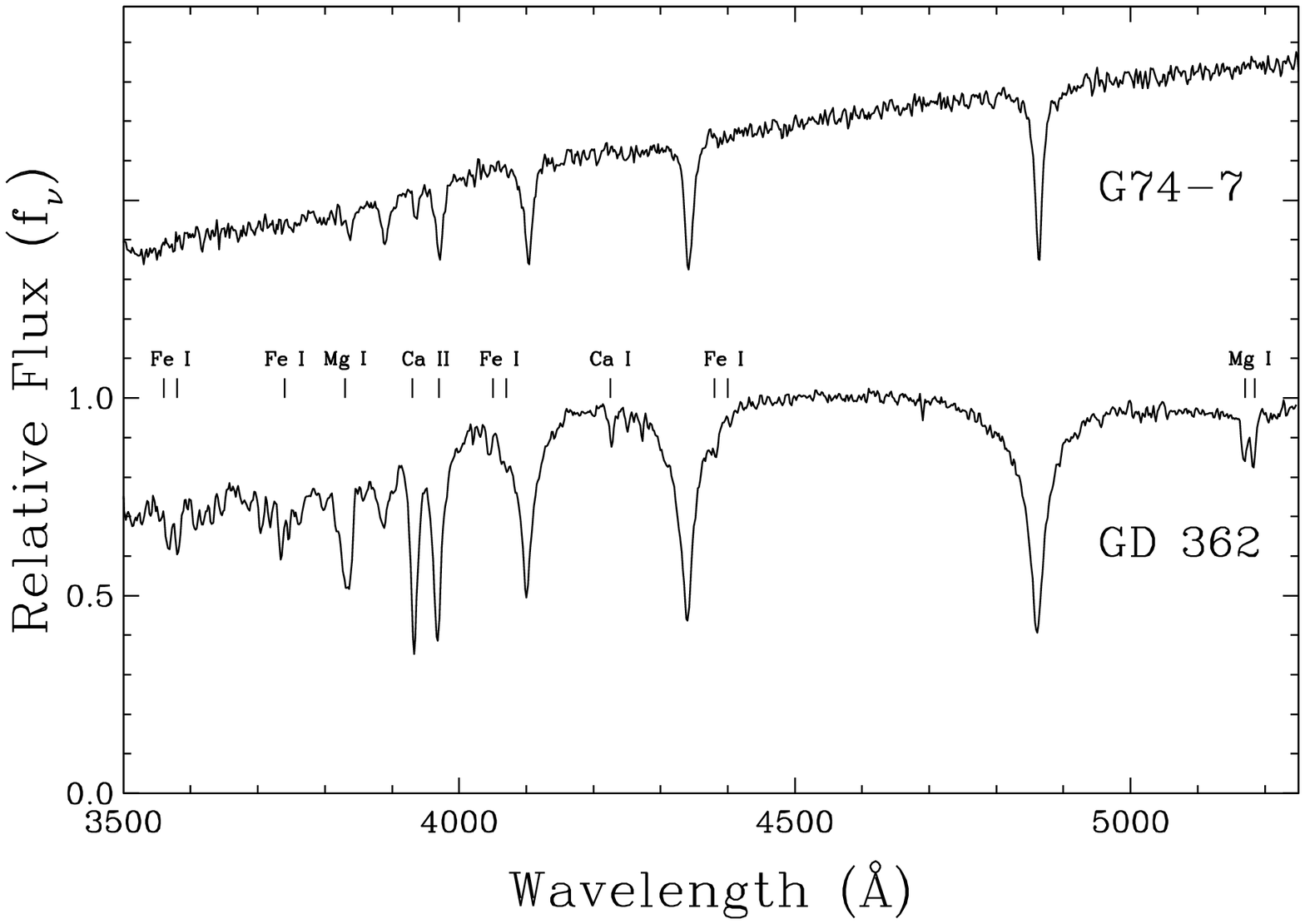] {Flux calibrated optical spectra of G74-7 ({\it top}) 
and GD 362 ({\it bottom}), normalized to unity
at 4500~\AA, and offset from each other by a factor of 0.7. The major
metallic features in the spectrum of GD 362 are indicated by tick
marks. In both cases, H$\epsilon$ is blended with Ca~\textsc{ii}
$\lambda$3969.\label{fg:f1}}

\figcaption[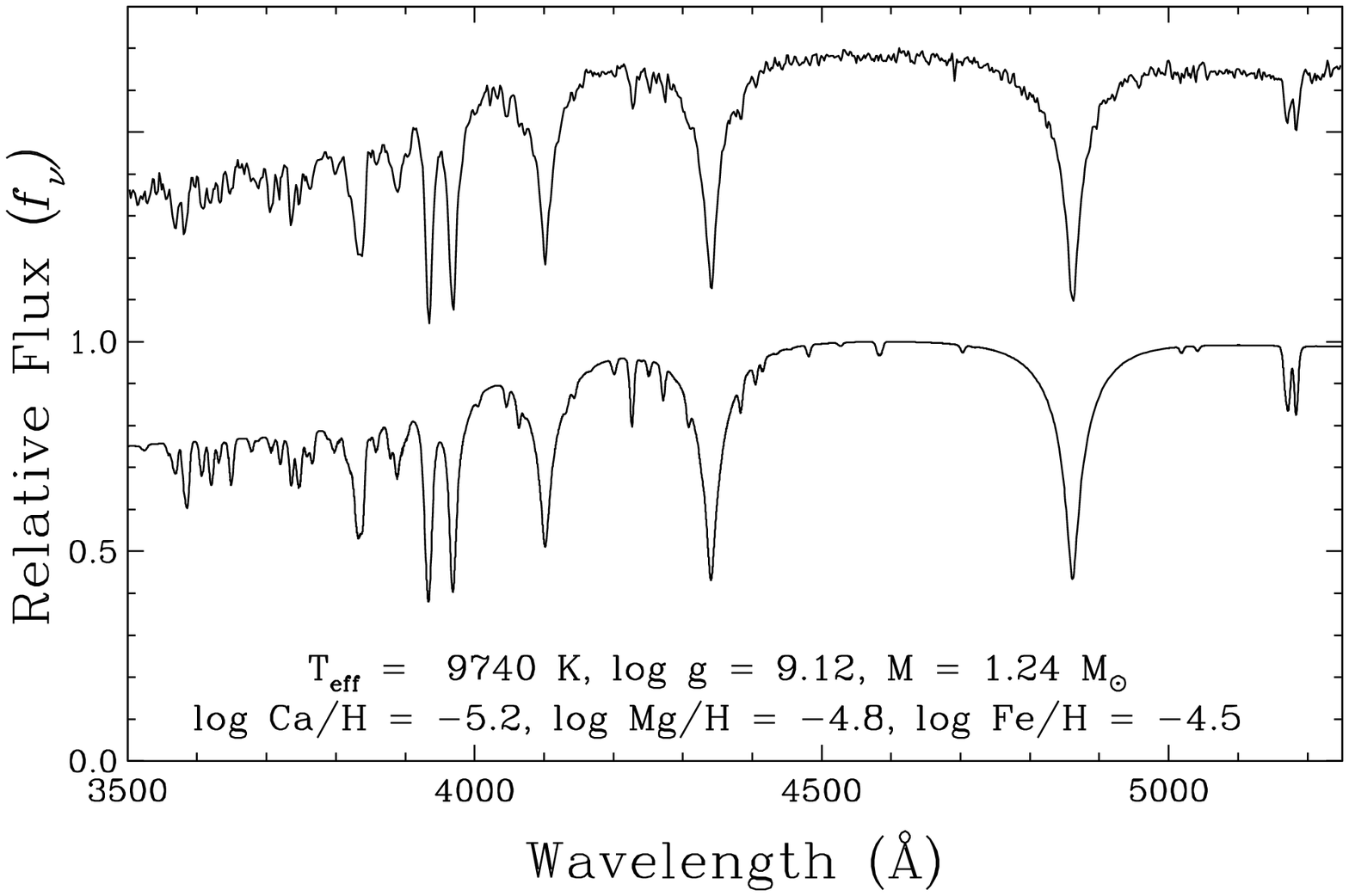] {Our best fit ({\it bottom}) to the observed spectrum 
of GD 362 ({\it top}). The atmospheric parameters obtained from this
analysis are indicated along with the mass derived from the
evolutionary models of \citet{fontaine01}. \label{fg:f2}}

\figcaption[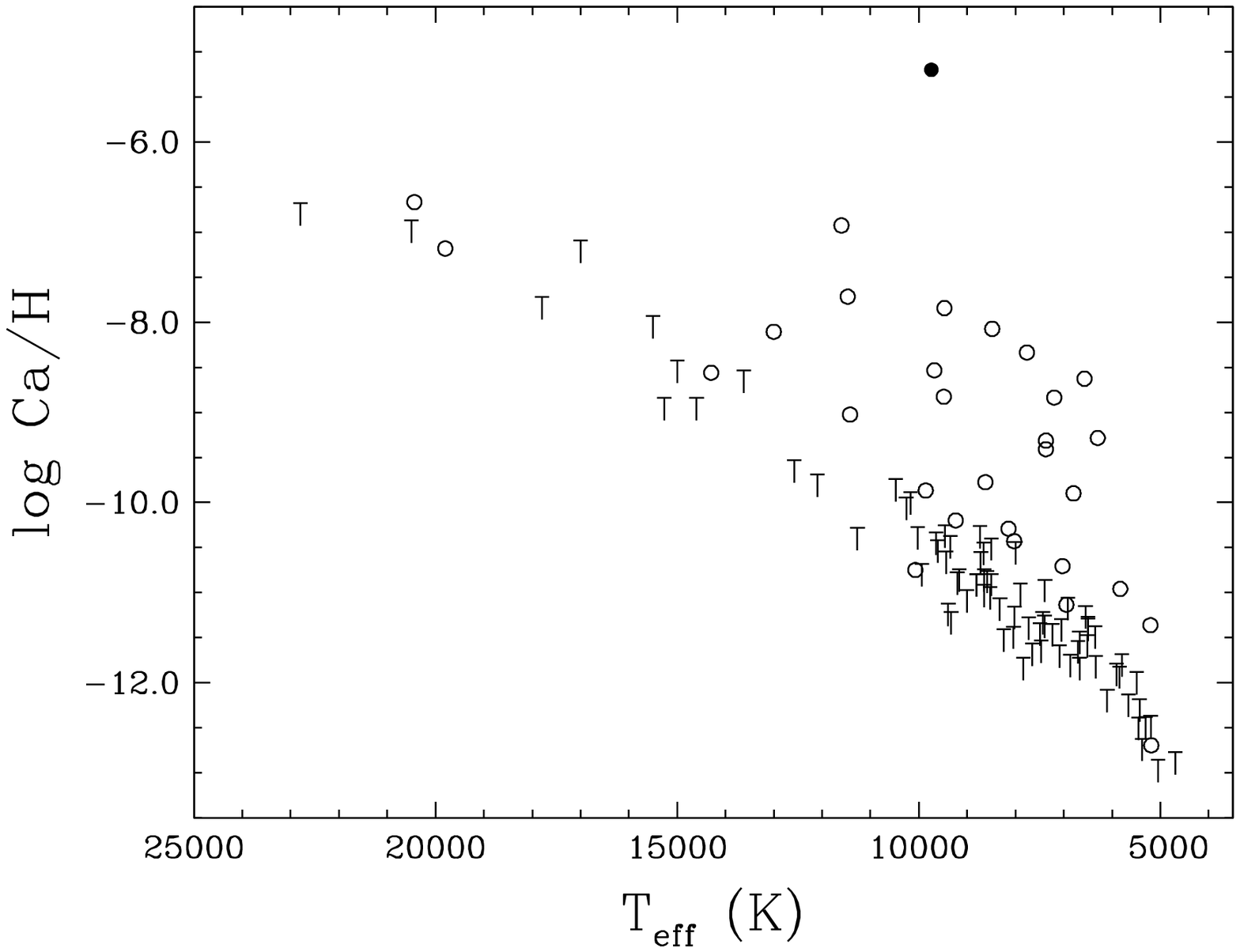] {Calcium abundances, or upper limits, as a function of
effective temperature for the white dwarfs taken from Tables 1 and 2
of \citet{zuckerman03}. The open circles mark objects where the Ca
\textsc{ii} K line was detected, while the crosses denote white dwarfs with
only upper limit determinations. The filled circle represents GD
362. \label{fg:f3}}

\clearpage

\begin{figure}[p]
\plotone{f1.eps}
\begin{flushright}
Figure \ref{fg:f1}
\end{flushright}
\end{figure}

\begin{figure}[p]
\plotone{f2.eps}
\begin{flushright}
Figure \ref{fg:f2}
\end{flushright}
\end{figure}

\begin{figure}[p]
\plotone{f3.eps}
\begin{flushright}
Figure \ref{fg:f3}
\end{flushright}
\end{figure}

\end{document}